\documentclass[a4paper]{article}
\def\hess{H.E.S.S.\ }
\def\degree{$^{\circ}$}
\usepackage{icrc2013}
\usepackage{color}
\usepackage[normalem]{ulem}
\usepackage{amsmath}

\title{Limits on Primordial Black Hole evaporation with the H.E.S.S. array of Cherenkov telescopes}

\shorttitle{H.E.S.S. limits on PBH evaporation}

\authors{
J-F.Glicenstein$^{1}$,
A.Barnacka$^{2}$,
M.Vivier$^{1}$,
T.Herr$^3$,
for the H.E.S.S. Collaboration.
}

\afiliations{
$^1$ CEA Saclay, DSM/Irfu, F-91191 Gif-Sur-Yvette Cedex, France \\
$^2$ Nicolaus Copernicus Astronomical Center, ul. Bartycka 18, 00-716 Warsaw, Poland \\
$^3$ Max-Planck-Institut f\"ur Kernphysik, P.O. Box 103980, D 69029 Heidelberg, Germany \\
}

\email{glicens@cea.fr}

\abstract{
Data collected by the \hess array between 2004 and 2012 have been used 
to search for photon bursts from primordial black hole explosions. Bursts were searched for in a 30 second time-window.  
The duration of the search window has been optimized to increase the burst signal while keeping the statistical background low. 
No evidence for a burst signal was found. 
Preliminary upper limits on the local rate of PBH explosions of $1.4\times 10^{4}\, \mathrm{pc}^{-3} \mathrm{yr}^{-1}$ have been obtained, which improve previously published limits by almost an order of magnitude.   

}


\begin{document}
\maketitle

\section{Introduction}
Primordial black holes (PBH) \cite{1974MNRAS.168..399C} are compact objects 
that may have been formed in the early Universe via a variety of mechanisms. These include (see e.g. \cite{2005astro.ph.11743C} for a review)
the gravitational collapse of overdense regions with significant density fluctuations, pressure reduction or bubble collisions during cosmic phase transitions, and collapse of topological defects such as cosmic strings or domain walls. The mass function of PBHs depends on the formation mechanism. PBHs could have masses ranging from 10$^{-5}$g for PBHs created at the Planck time \cite{2010PhRvD..81j4019C} upwards.     

Black holes 
were predicted by Hawking \cite{1974Natur.248...30H} to radiate off particles with a black body spectrum of energies. The emission can thus be described by an effective temperature
\begin{equation}
 T_{\mathrm{BH}} = \frac{M_p^{2}}{8\pi M_{\mathrm{BH}}},
\end{equation}
where $\mathrm{M_p}$ and $M_{\mathrm{BH}}$ are the Planck mass and the PBH mass respectively.
For black holes of stellar masses or higher, Hawking's radiation is quite negligible but for small enough PBHs, it becomes the predominant process that governs the black hole evolution. Black holes lose their mass by Hawking radiation at a rate inversely proportional to their squared mass:
\begin{equation}
 \frac{dM_{\mathrm{BH}}}{dt} = -\frac{\alpha(M_{\mathrm{BH}})}{{M_{\mathrm{BH}}}^{2}}, 
\end{equation}
where $\alpha(\mathrm{M_{\mathrm{BH}}})$ is a parameter counting the number of degrees of freedom available to the radiated particles. The parameter $\alpha$ is an increasing function of the black hole temperature and strongly depends on the particle physics model at high energies \cite{1991Natur.353..807H}. Since the particle emission rate increases with black hole temperature, PBH evaporation is a runaway process that eventually leads to a violent explosion and bursts of particles. PBHs can evaporate more or less rapidly depending on the number of available particle species that can be produced. In a Friedman universe and in the standard model of particle physics, PBHs whose initial mass does not exceed 5 $\times$ 10$^{14}$g are expected to have fully evaporated within the 10$^{10}$ years of our Universe history. Consequently, PBHs a little more massive than this will still be emitting particles at a rate large enough so that they would be detectable.\\
The cosmological constraints on PBHs have been reviewed 
by Carr et al \cite{2010PhRvD..81j4019C}. 
The best method for constraining low mass PBHs ($M_{\mathrm{BH}} \le 5\times 10^{14}$ g) is thus through their $\gamma$-ray emission. Previous searches have attempted to detect a diffuse photon signal from a distribution of PBHs \cite{2009A&A...502...37L} or to search directly for the final stage emission of an individual hole \cite{2006JCAP...01..013L,2008AstL...34..509P,2008AIPC.1085..701S,2012JPhCS.375e2024T}. 
The EGRET observation of the diffuse $\gamma$-ray background allowed to set an upper limit on the low mass PBH density $\Omega_{PBH}$ of $0.2\times 10^{-9}$ to $2.6\times 10^{-9}$ \cite{2009A&A...502...37L}. The search for direct PBH explosions through $\gamma$-ray bursts did not find any evidence of their presence yet. The current upper limits on the local PBH explosion rate lie in the 10$^{5}$-10$^{6}$~pc$^{-3}$ yr$^{-1}$ range \cite{2010PhRvD..81j4019C}. 
Note however that it has been argued that a class of very short gammay ray bursts are actually the final stage of PBH evaporation\cite{2011IJAA....1..164C}. \\

The present paper reports on the search for TeV $\gamma$-ray bursts with a timescale of a few seconds, as expected from the final stage of PBHs evaporation, using the H.E.S.S. array of Imaging Atmospheric Cherenkov Telescopes (IACTs). 
The H.E.S.S. array is presented in Sec. \ref{sec:hessarray}. A modelling of the expected PBH $\gamma$-ray signal is carried out in Sec. \ref{sec:modelling}. The data analysis procedure and the burst search strategy are presented in Sec. \ref{sec:dataanalysis} and \ref{sec:burstsearch}. Finally, preliminary upper limits on the local PBH explosion rate are derived in Sec. \ref{sec:results}.

\section{The H.E.S.S. array}\label{sec:hessarray}

H.E.S.S. is an array of five imaging atmospheric Cherenkov telescopes dedicated to observing very-high energy (VHE) $\gamma$-rays with energies above $50$~GeV 
from astrophysical sources. 
It is located in the Khomas Highland of Namibia. 
The first four telescopes have been installed in 2003 (H.E.S.S-1 phase of the experiment, with an energy threshold of $\sim$100~GeV) and have been operational since 2004. Each telescope of H.E.S.S-1 comprise a tesselated optical reflector of $107$m$^{2}$ \cite{2003APh....20..111B} and a camera with 960 photomultiplier tubes. The camera field of view is $5^{\circ}$ in diameter. 
The stereoscopic technique \cite{2004APh....22..285F} allows for an
accurate reconstruction of the direction and the energy of the primary 
gamma-ray. H.E.S.S-1 has an angular resolution of less than $0.1$\degree, 
a source location accuracy of   $\sim 30''$ for strong sources and an effective detection area of $\sim 10^{5}\,$m$^2$.
The sensitivity for point-like sources reaches $2\times 10^{-13}\,$cm$^{2}\,$s$^{-1}$ above 1 TeV for a 5$\sigma$ detection in 25 hours of a source at
a $20^{\circ}$ zenith angle \cite{2009APh....32..231D}.
A fifth telescope with a reflective area of $596\,$m$^2$ and a camera of 2048 photo multipliers has started its operations in 2012. This paper uses only data collected with the four telescopes of H.E.S.S-1.  

\section{Predictions for the PBH evaporation signal}\label{sec:modelling}

The theoretical number of $\gamma$-rays emitted from a PBH located at a distance $r$ and in the direction ($\mathrm{\alpha}$,$\mathrm{\delta}$) in the sky, during the last $\mathrm{\Delta}$t seconds of its life is given by:
\begin{equation}
N_{\mathrm{\gamma}}(r,\mathrm{\alpha},\mathrm{\delta},\mathrm{\Delta}t) = \frac{1}{4\mathrm{\pi}r^{2}} \int_{0}^{\mathrm{\Delta}t}dt \int_{0}^{\infty} dE_{\mathrm{\gamma}} \frac{d^{2}N}{dE_{\mathrm{\gamma}}dt}(E_{\mathrm{\gamma}},t) A(E_{\mathrm{\gamma}},\mathrm{\alpha},\mathrm{\delta}),
\end{equation}
where d$^{2}$N/dE$_{\mathrm{\gamma}}$dt is the instantaneous $\gamma$-ray spectrum emitted by the PBH at a time $t$ before complete 
evaporation. This spectrum is folded with the H.E.S.S. acceptance A(E$_{\mathrm{\gamma}}$,$\mathrm{\alpha},\mathrm{\delta}$) to take into 
account the instrument's efficiency in collecting $\gamma$-rays of energy E$_{\mathrm{\gamma}}$ at equatorial coordinates $(\mathrm{\alpha},\mathrm{\delta})$ in the sky. The response of the H.E.S.S. instrument to $\gamma$ rays depends on the 
zenith angle and offset angle of observation.  
The acceptance $ A(E_{\mathrm{\gamma}},\mathrm{\alpha},\mathrm{\delta})$ is an average over many runs with different zenith and offset angle\textbf{s}. 
It can be approximately factored into an energy dependent term  and a spatially dependent  term by writing
$ A(E_{\mathrm{\gamma}},\mathrm{\alpha},\mathrm{\delta})=A_{(0)}(E_{\mathrm{\gamma}}) A_{(1)}(\mathrm{\alpha},\mathrm{\delta}).$ 
The factor $A_{(1)}(\mathrm{\alpha},\mathrm{\delta})$ corresponds basically to the normalised sky acceptance map to $\gamma$-rays, which is maximum at the center of the camera and drops toward the edges. It is directly estimated from the data.   
The $A_{(0)}$ term is the effective area and is obtained from Monte Carlo simulations with different zenith angles and offsets. 

The  instantaneous $\gamma$-ray spectrum d$^{2}$N/dE$_{\mathrm{\gamma}}$dt emitted by the PBH  depends on specific particle physics models \cite{1991Natur.353..807H}. It is assumed in this paper that the standard model of particle 
physics remains valid at high ($>$~200~GeV) temperatures. 
The presence of an atmosphere around the PBH is an important theoretical question which is
still debated \cite{2008PhRvD..78f4043M}. The PBH atmosphere would drastically alter the evaporation signal \cite{1997PhRvD..55..480H,2002PhRvD..65f4028D} by suppressing the high energy component. In this paper, we assume that any existing PBH atmosphere has a negligible effect on the evaporation signal.

\begin{figure}[htb]
\centering{\includegraphics[scale=0.4, clip=true]{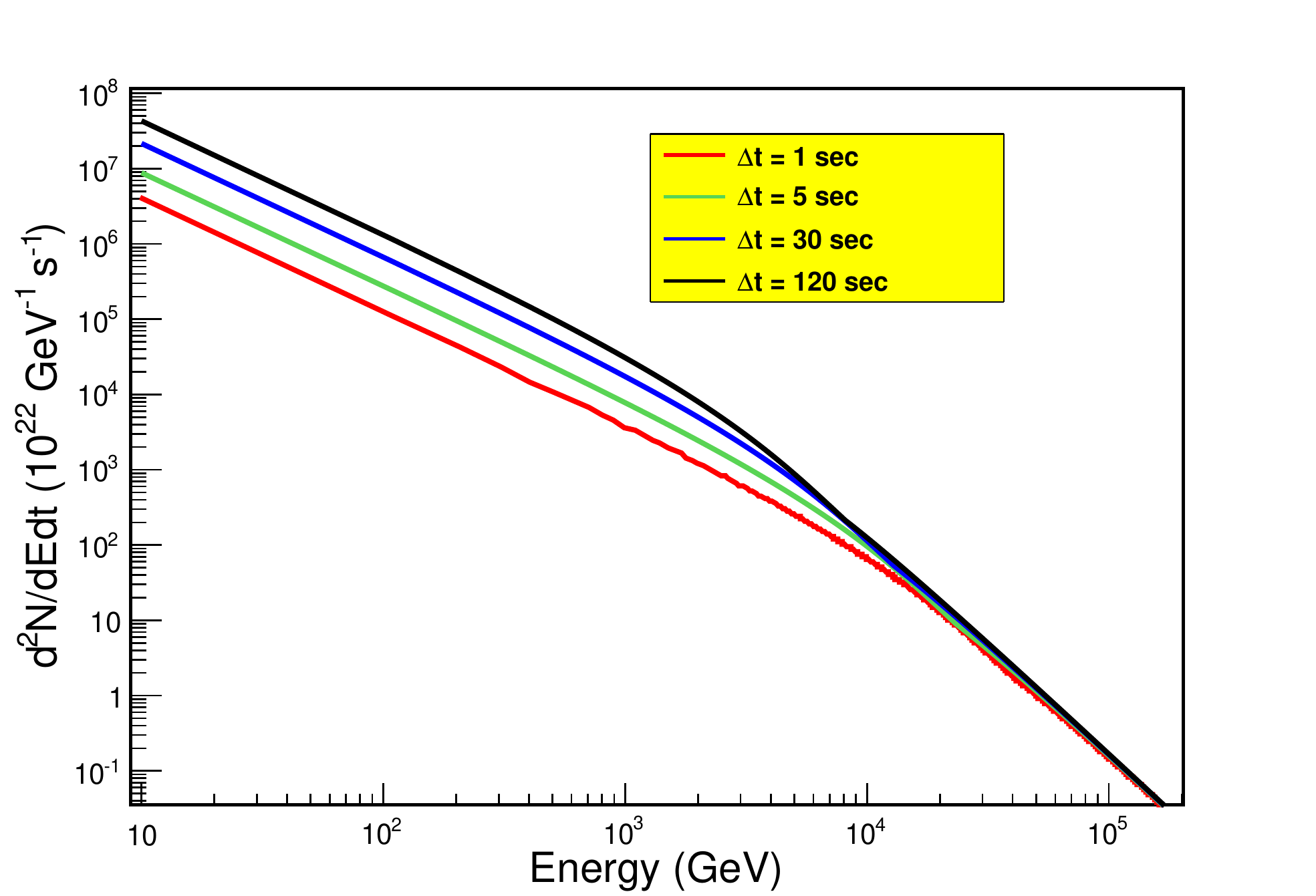}}
\caption{Integrated PBH spectrum at several values of the remaining time before explosion $\Delta$t.}
\label{fig:icrc2013-0930-01}
\end{figure}

The integrated spectrum above the energy $E_{D}$ is given by 
by Halzen et al \cite{1991Natur.353..807H} and shown on Fig \ref{fig:icrc2013-0930-01} for several values of the time remaining before total evaporation $\Delta$t. 
  
The signature of a PBH explosion consists in the detection of several photons within a time-window of a few seconds. The number of photons in the burst is the size of the burst.

The probability of detecting a burst of size $b$ when observing a PBH which emits N$_{\mathrm{\gamma}}$(r,$\mathrm{\delta}$,$\mathrm{\alpha}$,$\mathrm{\Delta}$t) $\gamma$-rays follows a Poisson statistics:
\begin{equation}
P(b,N_{\mathrm{\gamma}}) = e^{-N_{\mathrm{\gamma}}} \frac{N_{\mathrm{\gamma}}^{b}}{b!}
\end{equation}
Integrating this probability over space, and summing over each run give the number of expected bursts of size $b$ to be detected in the data:
\begin{equation}
n_{signal}(b,\mathrm{\Delta}t) = \mathrm{\dot{\rho}_{PBH}}V_{\mbox{eff}}(b,\mathrm{\Delta}t)
\label{eq:ngammath}
\end{equation} 
where $\dot{\rho}_\mathrm{PBH}$ is the local PBH explosion rate and the effective space-time volume of PBH detection is defined by 
\begin{equation}
V_{\mbox{eff}}(b,\mathrm{\Delta}t) =
\sum_{i} T_{i} \int d\mathrm{\Omega_{i}} \int_{0}^{\infty} dr r^{2} P_{i}(b,N_{\mathrm{\gamma}}),
\label{eq:effectivevolume}
\end{equation}
where the indice $i$ goes over each run of the H.E.S.S. dataset, T$_i$ and d$\mathrm{\Omega_i}$ being the corresponding run live time and observation solid angle respectively.\\

\begin{figure}[htb]
\centering{\includegraphics[scale=0.4, clip=true]{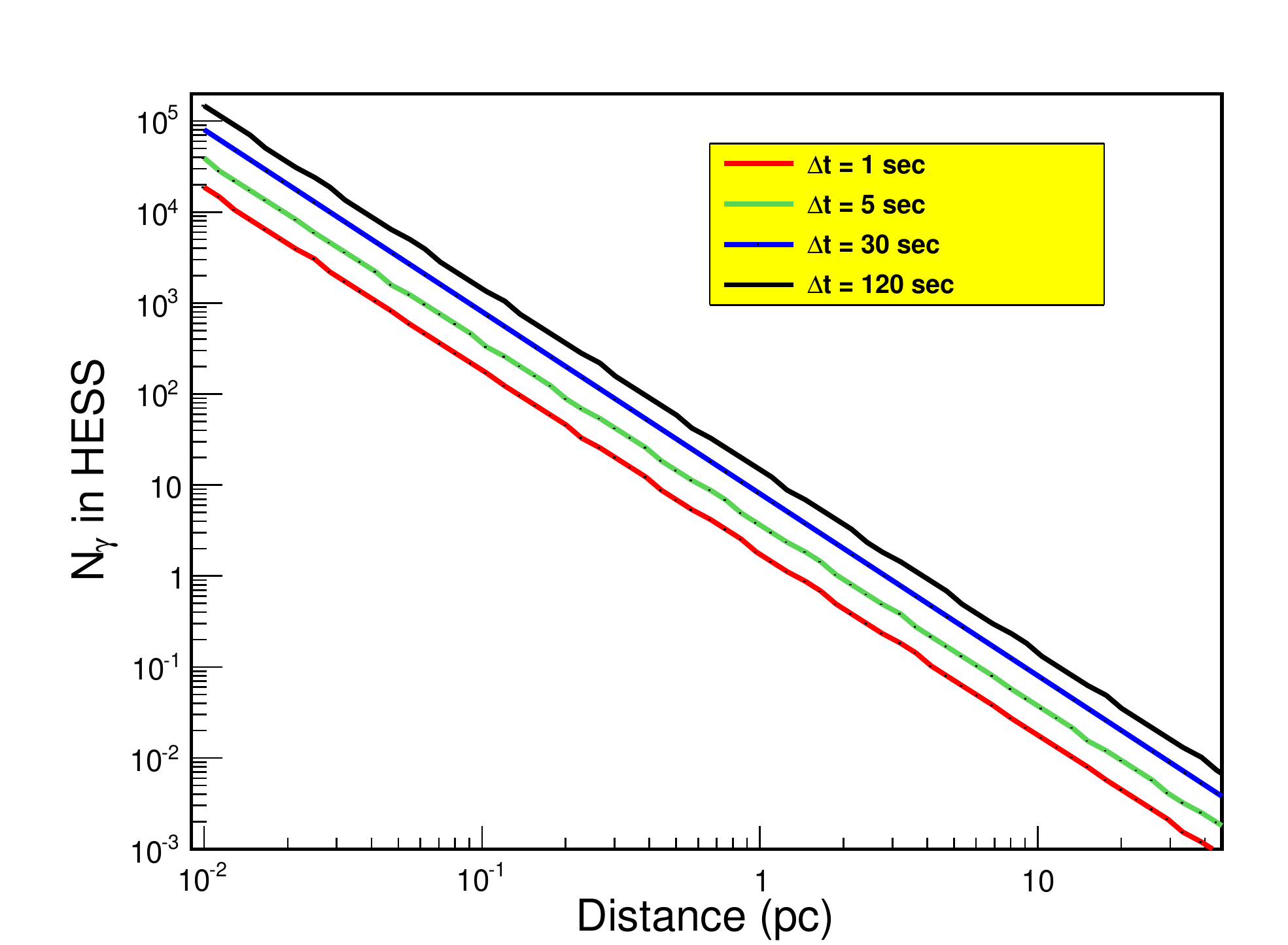}}
\caption{Average number of photons from a PBH explosion detectable by H.E.S.S. as a function of the distance to the burst, for several values of the remaining time before explosion $\Delta$t.}
\label{fig:icrc2013-0930-02}
\end{figure}

The average number of photon detected by H.E.S.S. as a function of distance is shown in Fig. \ref{fig:icrc2013-0930-02}.

The total number of bursts $n_\mathrm{tot}$ is the sum of the signal from PBH explosions $n_\mathrm{sig}$, given by equation (\ref{eq:ngammath}) and a contribution from statistical background bursts $n_\mathrm{back}.$  
A likelihood analysis is performed on the observed $n_\mathrm{obs}(b)$ bursts of size b$\geq$2 to find the optimal PBH explosion rate $\dot{\rho}_\mathrm{PBH}$.

\section{Data processing}\label{sec:dataanalysis}

\subsection{Data set selection}
The data set used for the burst search includes a large \\
fraction of  targeted and survey observations made with the H.E.S.S-1 array
 from March 2004 through May 2012. 
Observations are organized in runs of approximately $28$~min duration.
Poisson fluctuations of photon candidates (``$gamma$-like events'') 
arrival times could accidentally 
mimic PBH bursts. Since the major fraction of the H.E.S.S. observations is
taken towards astrophysical sources, the contribution of these sources
to background bursts has to be considered.
A few thousands photon candidates, mostly misidentified hadrons, are 
reconstructed in a typical observation run of an empty field. 
The actual number depends on the data taking condition and the reconstruction method. The candidate photon background in a point source search is thus a 
few photons per minutes, at the level of the count rate of the Crab nebula. 
Thus, strong sources such as the active galaxies PKS~2155-304, MRK~421 and the Crab pulsar wind nebula have to be excluded. However, most VHE gamma-ray sources are too weak to accidentally mimic PBH bursts.
The evaluation of the effective sensitivity volume of observations (equation
\ref{eq:effectivevolume}) implies the calculation of spatial acceptance maps. 
To obtain maps with sufficient statistical accuracy,  
targets with only a small observation time were excluded from the data set.
Only observations where at least three of the four telescopes 
participated in data taking were used to improve the angular resolution. 
Finally, each run has to pass certain quality criteria which 
ensure that the data used for analysis was taken under good environmental and instrumental conditions. After these cuts
\sout{cuts}, the data set includes 6424 runs, covering $\sim 45\%$ of all H.E.S.S-1 observations between 2004 and 2012, and corresponds to roughly 2600~hours of 
observation time. 

\subsection{Data reduction}
The vast majority of imaged air showers in the data is not caused by VHE $\gamma$-rays but by an unwanted background of 
hadronic cosmic rays.
To suppress this background and reconstruct the direction and energy of 
the $\gamma$-ray candidates, an implementation of the 'model' technique\cite{2009APh....32..231D}, Model++ has been used. In the 'model' technique, 
the air showers are described by a semi-analytical model. Expected properties of the camera images are then compared to the observational data based on a maximum likelihood method. The model technique is known to provide an improved sensitivity, particularly at lower energies, and a  better hadron rejection compared to the more traditionnal Hillas reconstruction.
The angular resolution, defined as the 68\% containment radius, is 0.06\degree. 
 Additional cuts such as a minimum charge of 60 photo-electrons were applied.  
Gamma-like events with a distance to the center of the camera larger than 2$^\circ$ are excluded. 
The analysis chain and the algorithms have been cross-checked with an independent analysis chain.
Finally, arrival times and geometrically reconstructed arrival directions in the RA-Dec (J2000) coordinate system of all $\gamma$-like events (a few thousands per run) were stored in event lists.  

\section{Searching for TeV $\gamma$-ray bursts}\label{sec:burstsearch}
\subsection{The burst search algorithm}
The event lists have been searched for bursts of different durations $\tau = 1$, $5$, $10,$ $30,$ $45,$ $60$ and $120$ seconds. As these time scales are much 
shorter than the duration of a single run, all runs can be analyzed individually. Each of the $N_{ev}$ entries $i$ stored in the 
run's event list marks the start time $t_i$ of a possible burst that could include additional $\gamma$-like 
events reconstructed within the time interval  $[t_i, t_i+\tau]$.
Since PBHs are point sources, these additional $\gamma$ candidates
were searched in a circle of radius $\theta = 0.1$\degree  in the RA-DEC plane.
This radius corresponds to a 90\% containment probability for point sources.
As the sensitivity within the H.E.S.S. FoV for VHE gamma-ray events drops off rapidly for angular distances 
greater than approximately $2^\circ$ from the telescopes pointing direction, the burst search is restricted to the inner $2^\circ$ of 
the FoV. To account for the finite size of PSF we include all events within a maximum distance 
of $2.1^{\circ}$ to the telescopes's pointing direction. For all events lying within the time interval  $[t_i, t_i+\tau],$ 
the burst search algorithm finds the maximal subset that fits in a circle with radius $\theta$ in the RA-Dec plane. 
This maximal subset is said to be a burst of a size $b.$ 
Each photon candidate is thus associated with a burst size $b.$ 
To prevent multiple counting of bursts, the number $N(b)$ of detected bursts of size $b$ is defined as the number of events 
$N_{ev}(b)$ that have been assigned the burst size $b$ divided by $b$ \cite{2006JCAP...01..013L}:
\begin{equation}
  N(b) = \frac{N_{ev}(b)}{b}
\end{equation}
Using this convention, the following intuitive normalization relation holds:
\begin{equation}
  \displaystyle
  N_{ev} = \sum_{b}N_{ev}(b) = \sum_{b} b N(b)
\end{equation}
Note that the maximal subset defining $b$ is not necessarily unique, as there may be more than one valid maximal subset. Also, by optimizing for the largest possible burst size, the algorithm may underestimate the number of smaller size bursts. However, because the number of bursts $N(b)$ found at burst size $b$ is always much greater than $N(b+1)$ this effect can safely be neglected.

\subsection{Background estimation}\label{sec:background}
\begin{figure}[htb]
\centering{\includegraphics[scale=0.4, clip=true]{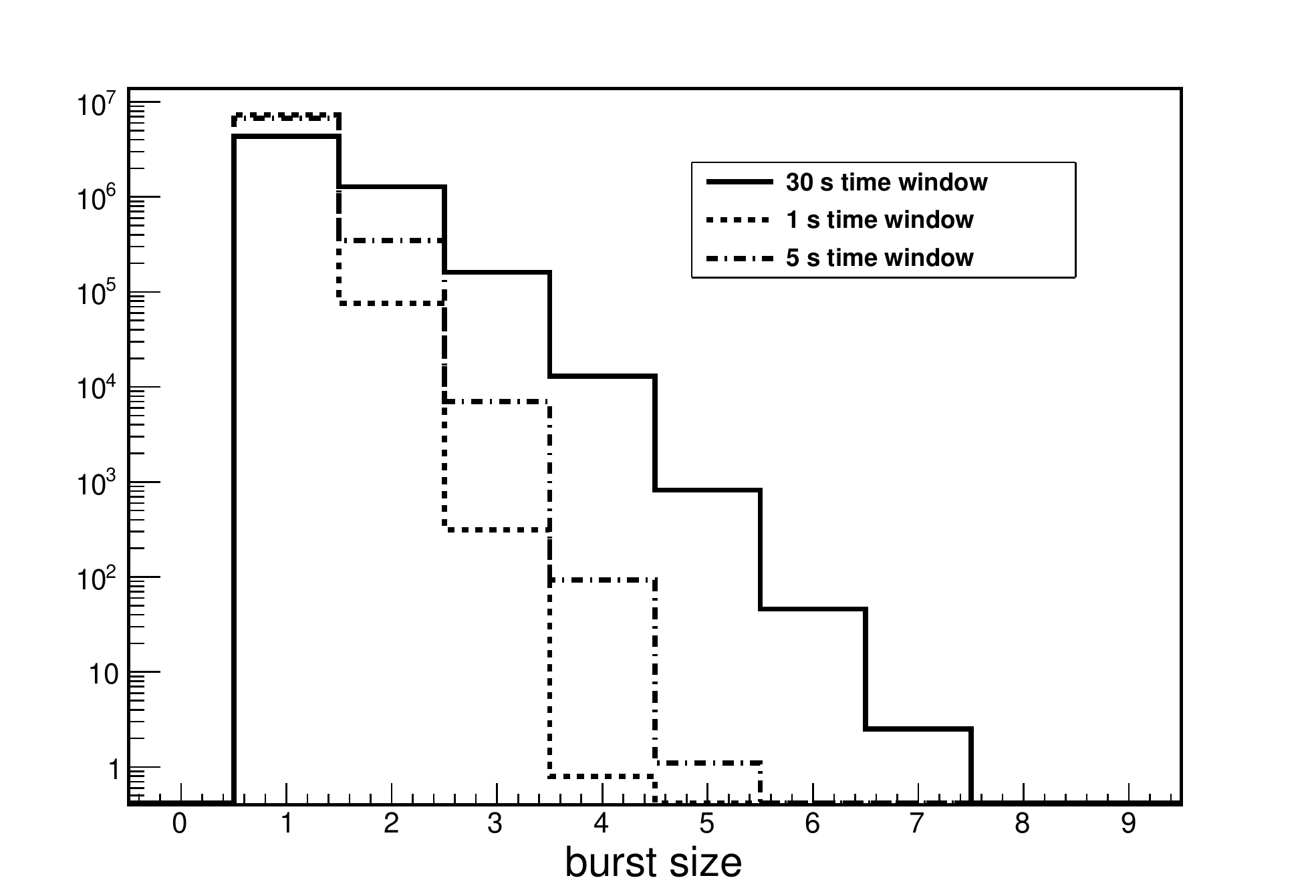}}
\caption{Simulated background burst size spectrum for bursts searches  in the H.E.S.S. dataset within different $\tau$ search time windows.}
\label{fig:backcomp}
\end{figure}

The major background to 
physical bursts is caused by cosmic ray primaries that accidentally happen to arrive from neighboring directions within a narrow time window. Estimating the contribution of this statistical background is essential for extracting a possible VHE gamma-ray burst signal. In our analysis, the estimation
of background relies  on the ``scrambling'' method. 
In the scrambling method, new simulated datasets are created by keeping the arrival direction of each $\gamma$ candidate and scrambling their arrival times.
This method automatically accounts for all instrumental characteristics and effects determining the spatial distribution of events in the FoV. 
To reduce statistical errors of the background estimate, ten simulated datasets were produced for each H.E.S.S. observation run.\\
The simulated background burst size spectrum, calculated for the whole H.E.S.S. dataset, is shown on Fig. \ref{fig:backcomp} for several values of the search time window $\tau.$

\subsection{Time-window selection}\label{sec:timewindowselec}
Previous PBH explosion searches with Cherenkov telescopes \cite{2006JCAP...01..013L,2012JPhCS.375e2024T} have used a $\tau = 1$ s time-window.
However, the optimal time-window to search for PBH explosions is given by a compromise between the effective volume and the statistical background, which both increase with $\tau$, albeit at different rates. The optimal time-window
obviously depends strongly on the rejection of hadrons by the photon analysis program. Since Model++ has a very good rejection of hadrons, it turns out that it is possible to extend the time-window $\tau.$ 
The optimal time-window was selected by computing a sensitivity limit with 250 hours on the field of view of radiogalaxy Centarus A. The sensitivity
limit is obtained by optimizing the likelihood $L$ under the assumption 
that $n_\mathrm{obs}(b)=n_\mathrm{back}(b).$ The variation
of the sensitivity limit as a function of the duration of the search window $\tau$ 
has a broad minimum around 
$\tau=30$s. This value of $\tau$ was used for the limits on the PBH evaporation rate derived in section \ref{sec:results}.

\section{Results and discussion}\label{sec:results}

\begin{figure}[htb]
\centering{\includegraphics[scale=0.4, clip=true]{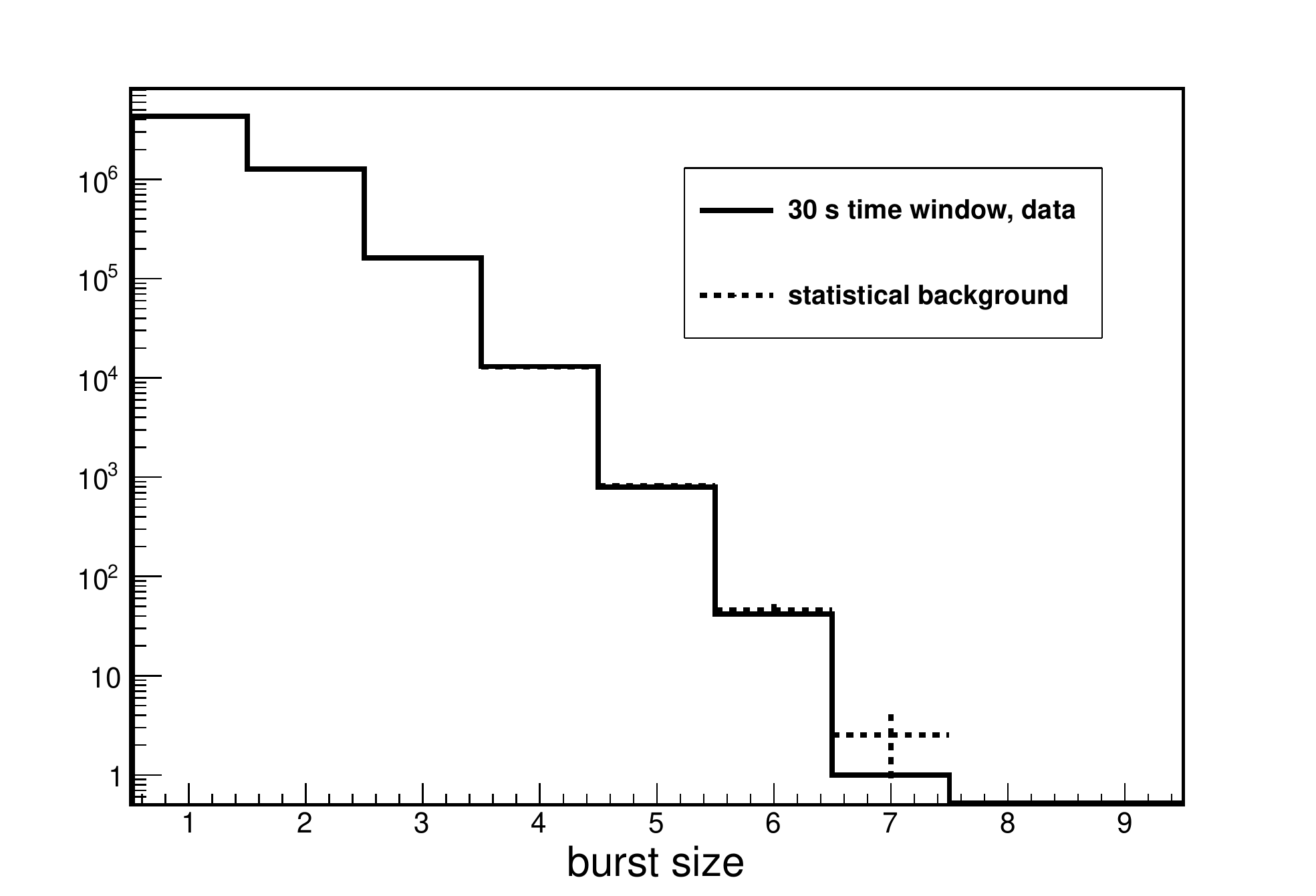}
}
\caption{Preliminary observed burst size distributions (solid line) and estimated statistical background (dashed line). The search time window has the nominal value $\tau = 30 s$. 
} \label{fig:size30s}
\end{figure}

Bursts were searched for in the dataset defined in Sec \ref{sec:dataanalysis}. The observed burst size spectrum is shown by a solid line on 
Fig. \ref{fig:size30s} for the nominal value of the search window $\tau = 30$ s. The estimated statistical background is displayed with a dash line. 
Fitting the observed size spectrum by the expected statistical background gives
$\chi^2/\mbox{d.o.f} = 4.2/7$. This shows that the observed spectrum is in excellent agreement with the statistical background.


The H.E.S.S. dataset shows thus no indication for an excess of bursts over the statistical background.  

The agreement between the observed burst distribution and the estimated 
statistical background can be translated into an upper limit on the PBH
explosion rate $\dot{\rho}_\mathrm{PBH}.$ 
The 95\% CL upper limit on $\dot{\rho}_\mathrm{PBH}$ is obtained 
by demanding that $2\ln{L} < 3.84.$ 

The preliminary upper limit on the explosion rate is 
$\dot{\rho}_\mathrm{PBH} < 1.4\times 10^4 \mbox{pc}^{-3}\mbox{yr}^{-1}$ at the 95\% CL 
for $\tau =30\,$s. The sensitivity limit, defined in section \ref{sec:timewindowselec} is $1.7\times10^4 \mbox{pc}^{-3}\mbox{yr}^{-1}.$  By comparison, the preliminary
upper  
limit obtained with the $\tau = 1\,$s search time-window is
$\dot{\rho}_\mathrm{PBH} < 4.9\times 10^4 \mbox{pc}^{-3}\mbox{yr}^{-1}$ (95\% CL).
 
The 95\% upper limit on the local rate of PBH explosion obtained with the $\tau = 30 s$ search time-window improves the best published result obtained with Cherenkov telescope arrays \cite{2012JPhCS.375e2024T} by almost an order of magnitude.  
Improvement of a factor of $10^4$ is still needed before the hypothesis that the very short gamma ray bursts originate from PBH explosions \cite{2011IJAA....1..164C} can be tested.  Our limit depends
strongly on the hypothesis that the PBH atmospheric effects are negligible. In the second phase of H.E.S.S, H.E.S.S-2, it will become possible to 
constrain models of PBH with atmospheres such as that of Dahigh and Kapusta \cite{2002PhRvD..65f4028D} thanks to the lowering of the energy threshold below 50 GeV.

\footnotesize{{\bf Acknowledgment:}{
Please see standard acknowledgement in H.E.S.S. papers, not reproduced here due to lack of space.
}}

\end{document}